\documentclass[aps,preprint,epsfig,rotate]{revtex4}
\usepackage{graphicx}
\usepackage{bm}
\usepackage{epsfig}
\usepackage{amsmath}

\newcommand{\be}{\begin{equation}}
\newcommand{\ee}{\end{equation}}
\newcommand{\bea}{\begin{eqnarray}}
\newcommand{\eea}{\end{eqnarray}}

\begin{document}

\title{Accurate CI and Hylleraas-CI wave functions for the atomic effects in the whole-atom-nuclear $\beta^-$-decay of the Li atom}

\author{Mar\'{\i}a Bel\'{e}n Ruiz}
\email[E--mail address: ]{maria.belen.ruiz@fau.de}                   

\affiliation{Department of Theoretical Chemistry\\
Friedrich-Alexander-University Erlangen-N\"urnberg,
Egerlandstra\ss e 3, 91058 Erlangen, Germany}

\date{\today}

\begin{abstract}
In this review we study the level of accuracy of the electronic wave functions which
is necessary to describe properly the atomic effects during nuclear $\beta$%
-decay. In the case of the $\beta^ -$-decay in the Li atom into Be$^+$ ion
we compare the numerical values of the transition probabilities from the  
$S$, $P$, $D$ low-lying states of the  
initial atom and final ion calculated using both Hylleraas-Configuration
Interaction (Hy-CI) and Configuration Interaction (CI) with Slater orbitals
wave functions. In addition using the CI method the transition probabilities from 
$F$, $G$, $H$ and $I$ low-lying states have been calculated.   
The average of the absolute deviation of the transition probabilities
distribution for low-lying $S$ states is $<0.15\%$, for $P$ states $<0.5\%$, and
larger for D- and higher energy states. The numerical results demonstrate
that for low-lying states the atomic effect parameters in $\beta$-decay can
be calculated with sufficient accuracy using CI wave functions constructed
with Slater orbitals. This result opens a new avenue for the accurate
calculation of atomic effects during the $\beta$-decay in heavier atoms and
molecules.
\end{abstract}


\maketitle

\newpage

\section{Introduction}

In this review we study the redistribution of the electronic density in
nuclear $\beta $-decay processes in atoms. This is a new field lying in the
frontiers of nuclear physics, atomic/molecular physics and quantum
chemistry. In this paper we study the excitations, ionization processes and
electronic rearrangement during nuclear reactions of $\beta$-decay employing
quantum mechanical methods. In this work we approach the phenomenon of
$\beta$-decay from the point of view of atomic and molecular physics and
chemistry. With this purpose, we introduce first some concepts of $\beta$-decay 
and the nuclear $\beta$-decay theory.

There are thousands of nuclides, from then only more than 300 are stable
nuclides. The rest of them experiment nuclear transformations. $\beta $%
-decay is a nuclear process in which a nucleus which is rich in neutrons or
protons decays emitting an electron or positron, a $\beta $-particle, in
order to become more stable. The phenomenon of $\beta $-decay occurs via the
week interaction force or quark-transformation.  
Neutrons and protons are built out of quarks. A proton consists on two
up-quarks ($u$-quark, charge $+\frac 23e$) and one down-quark ($d$-quark, charge 
$-\frac 13e$), while the neutron is built out of two $d$-quarks and one
$u$-quark. The mass of the proton is $1.00727647$ u and the mass of the
neutron $1.0086647$ u, and so the mass of the neutron is about 2$\%$ larger than the one of the proton.
In a neutron decay a $d$-quark is transformed into a $u$-quark and a virtual
boson $W$ (the virtual bosons take the charge and have a very short
life), which again produces an electron and an antineutrino $(d\rightarrow
u+W^{*-}\rightarrow u+e^{-}+\overline{\nu_e})$. In the proton decay a
$u$-quark transforms into a $d$-quark and a positive virtual boson, which decays
into a positron and a neutrino $(u\rightarrow d+W^{*+}\rightarrow
d+e^{+}+\nu _e)$. The neutral $Z^0$ bosons involved in interactions of neutrinos and 
matter, together with the $W^{+}$ and $W^{+}$ bosons 
are the intermediate particles for the weak interaction, which unified with
electromagnetism in the Standard Model of particle physics form the
electroweak force. For more details, see Refs. \cite{Halliday,Rohlf}.

In this paper we study the atomic effects for the so-called 'superallowed' 
$\beta$-transition. In the superallowed $\beta $-decays, the quantum
mechanical wave function of the entire nucleus does not change during the 
$\beta $-decay process except by the conversion between neutron and proton.
These decays are good tests of the Standard Model of describing electroweak
interactions because they are easier to be studied by using theoretical
methods and precise experimental measurements.

There are two main types of beta decay: $\beta ^{-}$-decay and $\beta ^{+}$%
-decay. In the $\beta ^{-}$-decay a neutron decays into a proton ($%
n\rightarrow p^{+}+e^{-}+\overline{\nu _e}$). The emitted $\beta ^{-}$%
-electron is leaving the nucleus with an speed close to the velocity of the
light. As a consequence, the nucleus is transformed ($_Z^AZ\rightarrow
_{Z+1}^AZ+e^{-}+\overline{\nu _e}$) into the nucleus of another element with
one charge more $Z+1$ and constant mass number $A$. The radioactive
Carbon-14 dating reaction is a good example of $\beta ^{-}$decay, i.e. $%
_6^{14}C\rightarrow _7^{14}N+e^{-}+\overline{\nu _e}$.

In the $\beta ^{+}$-decay a proton decays into a neutron ($p^{+}\rightarrow
n+e^{+}+\nu _e$). In this case, the nucleus transmutes into a new nucleus of
an element with one charge less ($_Z^AZ\rightarrow _{Z-1}^AZ+e^{+}+\nu _e$).
Note, that the mass of the neutron is larger than the mass of a proton,
therefore $\beta ^{+}$-decay occurs in nuclei with more than one proton,
since the missing mass is taken from the binding energy of the whole
nucleus. Example of $\beta ^{+}$-decay is the decay of the artificial
isotope of the fluor atom $_9^{18}F\rightarrow _8^{18}O+e^{+}+\nu _e$, which
is employed as radiotracer in nuclear medicine. From the numerical point of
view, $\beta ^{+}$-decay parameters are very difficult to be calculated,
since the final atoms are negatively charged, and the additional atomic
electron can be weakly bounded, being therefore difficult to determinate
these states using quantum mechanical methods.

There are other two types of $\beta $-decay. In the electron-capture, nucleus
which are rich in protons capture an electron of the $K$, or $L$ shell. The
electron vacancy is filled by another electron of the atom, which passes
to occupy a lower shell emitting a $X$-ray or in some cases when the atom is
in a excited states a $\gamma $-ray ($p^{+}+e^{-}\rightarrow n+\nu$). In the
bound-state $\beta $-decay, the emitted $\beta $-electron does not leave the
atom and it is kept in a bounded orbital.

In general, during $\beta $-decay and other nuclear processes the atoms are
not bare nucleus, they are surrounded by all or part of their 'atomic
electrons'. The change of the charge of the nucleus ($%
_Z^AZ\rightarrow _{Z+1}^AZ$ and $_Z^AZ\rightarrow _{Z-1}^AZ$) affects the electronic
states of the orbital electrons.

In the last years the development of the laser spectroscopy allows the meassurement of small  
changes in the spectra of atoms and molecules. These changes are produced by changes in the structure of the nucleus 
and by cooperative neutron-electron-gamma-nuclear processes including excitation, ionization, electronic rearrangement 
induced by nuclear reactions and $\beta$-decay \cite{Glushkov-spect,Yudin}. 
The same is true for muonic atoms, i.e. atoms with metastable nuclei surrounded by a negative muon ($\mu^-$), can 
experiment $\mu^-$-capture under emission of $\gamma$-ray. Muonic processes are also studied using 
atomic and nulear physics methods, are used as tools for nuclear spetroscopy and are 
good tests for the Standard Model \cite{Glushkov-mu}.     

In this work we study the transition probabilities of the transition from an
electronic state in  the inital atom to another electronic state in the
final atom in the case of the He and Li atoms using different quantum
mechanical methods. The transition probabilities to excited states have not
been yet evaluated for $>99\%$ of the atoms and ions. 

\section{The atomic theory of $\beta$-decay}

The decay is a reaction of kinetic first order, where the rate of the decay
is given by 


\begin{equation}
N(t)=N_0e^{-\lambda t}
\end{equation}
$\lambda $ is the ratio of decay or decay constant related to the half-life $%
t_{1/2}=\ln 2/\lambda =\tau \ln 2$, and the lifetime $\tau =1/\lambda $.
Similarly than in $\alpha $-decay, it is expected that the states of the
initial and final atom should play an important role in the decay constant $\lambda 
$. It is well-known that in nature the $t_{1/2}$ can variate from fractions of
seconds to millions of years. The first theory of nuclear $\beta $-decay was
proposed by Fermi in 1934 \cite{Fermi}. Fermi took the equation of the
emission of photons from excited states (note the use of the words 'initial'
and 'final') and used for the $\beta $-decay process. The equation is 
called 'Fermi's Golden Rule' \cite{Krane}:


\begin{equation}
\lambda =\frac{2\pi }\hbar \left| \int \Psi _{final}^{*}V_p\Psi
_{initial}d\tau \right| ^2\rho (E_f)=\frac{2\pi }\hbar \left| M_{f,i}\right|
^2\rho (E_f)
\end{equation}
where $\Psi $ is the total wave function of the complete system. In the first formulation 
the complete system includes  
nucleus and emitted particles ($\beta $-electron and neutrino). $M_{f,i}$ is
the matrix element between the initial and final wave functions. $V_p$ is
the operator of a very small perturbation that stimulates the transition. 
In the Fermi's theory of $\beta$-decay the form of $V_{p}$ was not known 
and the contribution of orbital electrons was not taken into account. Nowadays this potential
changes only the shape of the curve of the $\beta$-spectrum and is taken as


\begin{equation}
V_p\approx g\delta (r_n-r_p)\delta (r_n-r_e)\delta (r_n-r_{\overline{\nu }})%
\widehat{O}(n\rightarrow p)
\end{equation}
The $\Psi _{initial}$ is the wave function of the nucleus, whereas the $\Psi
_{final}$ includes also the $\beta $-electron and the neutrino. $\rho (E_f)$
is the density of final states determined usually with quantum statistics.
At Fermi's time it was still not known that the $\beta $-decay is produced via the 
weak interaction. It took about 20 years until the modern theory of $\beta$-decay 
was developed. The main equation in terms of the distribution of the
momentum $p$ of the $\beta $-electron $N(p)$ can be summarized as 
\cite{Krane}                      

\begin{equation}
N(p)=C_p^2(Q-KE_e)^2F(Z,p)\left| M_{f,i} \right| ^2S(p,q)
\end{equation}
$C_p^2(Q-KE_e)^2$ is an statistical factor of the density of final states of
the $\beta$-electron and neutrino. $F(Z,p)$ is the Fermi function, which
accounts for the Coulomb interaction $(eZ/r)$ between the charge of the
final nucleus and the charge of the leaving $\beta$-electron. $\left|
M_{f,i}\right| ^2$ is the matrix element between the initial and final
states (the same than in the Fermi's Golden Rule) and $S(p,q)$ a shape
factor correcting the matrix elements for the case of some forbidden $\beta$%
-decays.

The energy yielded during $\beta $-decay can be calculated from the binding
energy using the rest mass of the particles (in the decay of a neutron this
energy is about $0.78$ MeV). Energy and momentum are conserved during the
decay. In the decay from the ground state to a state $n$, $0 \rightarrow n$ \cite{Kaplan}


\begin{equation}
E^{(n)}_{\beta}=\Delta mc^2-m_e c^2-E_{\nu}-E_{rec}+\Delta E_{0n}
\end{equation}
$\Delta m$ is the mass defect in the nuclear transformation, $E_{\nu}$ is the energy 
of the neutrino, $E_{rec}$ is the recoil energy of the nucleus, and $\Delta E_{0n}$ is the difference 
between the electronic energies of the initial and final atom or molecule. 
As the final atom can be in different excited or ionized states $\Delta E_{0n}=\overline{\Delta E}$, 
the averaged electronic energy, which is  


\begin{equation}
\overline{\Delta E} = \sum_n w_{0n}\Delta E_{0n}
\end{equation}
where $w_{0n}$ are the transition probabilities for the transitions $0 \rightarrow n$ and they are 
calculated with the help of the sudden approximation, see next Sections


\begin{equation}
w_{f,i}=\left\vert \left\langle \Psi _{f}|\Psi_{i}\right\rangle \right\vert ^{2}
\end{equation}
where $f$ means final and $i$ initial states. Althought the total energy liberated during the
decay process is very large compared with the electronic excitation energies
of the atoms involved in the process, these excitation energies are
relevant in order to describe tiny effects in decay rates and half-times, to
describe chemical effects like bond-breaking and effect of the environment.
But the weight of the atomic transition probabilities $w_{f,i}$ is crucial in the
calculation of the neutrino mass in tritium $\beta $-decay in
order to obtain a reliable mass. The present neutrino upper mass limit from
experiments with tritium $\beta $-decay is $2eV/c^2$. Note that 
until now most of the investigations in $\beta $-decay including 'atomic
effects' were devoted to the case of tritium atom \cite{Brown} and tritium
molecule \cite{Kolos1,Kolos2}. Tritium atom is used for the determination of
the neutrino mass because it is the simplest system (apart of the bare
nucleus) with only one orbital electron and the calculation can be carried
out analytically.

The evaluation of the weights $w_{f,i}$ or transition probabilities requires
quantum mechanical calculations. The confidence in the exactness of the
quantum mechanical variational calculations is so high that disagreements
between experiment and theory required the repetition of experiments as it was the case i.e. of the
ground-ground transition in the HT molecule calculated by Wolniewicz 
($W_{0,0}=81.2\%$ ) \cite{Wolniewicz} and the experimental value $93,2(19)\%$.
In nuclear physics, except for the case of the neutrino, the matrix element 
$\left\vert M_{f,i}\right\vert ^{2}$ of Eqs. (2,4) has been mostly determined
using only nuclear wave functions ('bare nucleus' transition matrix
element), being the atomic effects added afterwards as corrections.

In the next sections we explain the importance of the atomic effects and
we present a method to calculate these effects accurately.

\subsection{Atomic effects in the whole-atom-nuclear $\beta $-decay}

Which physical effects are induced through 'atomic electrons'? And are these
'atomic effects' important for the 'whole-atom-nuclear' $\beta$-decay? In
the next we shall call 'atomic effects' to the effects produced by the
orbital electrons in nuclear $\beta$-decay. Here we shall enumerate them;
i) The change of the electron wave function due to the change of the atomic
electric field. This is the most important atomic effect, which we study by
using the 'sudden approximation', see next Section; ii) The change of the
integration limits in the calculation of the Fermi integral function because
the final and initial atoms are different systems; iii) The leaving 
$\beta$-electron is screened from the charge of the nucleus. The screening atomic
effect is taken into account approximatelly in the corrected Fermi function;
iv) A tiny energetic effect produced by the excitations in the final atom is
crucial to explain the shape of the end of the $\beta$-spectra, which is
employed for the determination of the neutrino mass; v) Exchange processes
involving the bound and continuum electrons (electromagnetic interactions
among the $\beta$-electron and the orbital electrons) are considered to be
very small. These exchange effects should be taken into account only in case
of comparison with very precise experiments; vi) Nuclear recoil, i.e. the
movement of the nucleus is the opposite direction as consequence of the
movement of the $\beta$-electron, is negligible in the case of the 
$\beta$-decay nuclear processes because the small mass of the electron; vii) The
phenomenon of 'double ionization' during $\beta $-decay with a half life
estimated in $10^{20}$ years is therefore very difficult to be detected
experimentally. The double decay can be produced by the successive 
$\beta$-decay of two neutrons or by an additional ionization during $\beta$-decay;
viii) In electron-capture $\beta $-decay the probability of the decay
life-times depend directly on the electronic density near of the nucleus 
\cite{Darmstadt1}. The vacancies created after electron capture are filled
with other bound electrons, producing 'vacancy cascades', the Auger Effect;
ix) In bound state $\beta$-decay the life-times depend on the number of
bound electrons, since these electrons screen the $\beta$-electron from the
nuclear charge. Bound state $\beta$-decay does not occur in neutral atoms,
since the electron will be in the continuum. In the practice bound-state 
$\beta$-decay occurs in heavily charged atoms in storage rings, where
dramatic changes of the nuclear life-times have been observed experimentally 
\cite{Darmstadt2}; x) 'Chemical effects' like the change of the valence
shell occupation numbers in molecules whose atoms experiment $\beta$-decay
processes can lead to different oxidation states between the initial and
final atoms and therefore to bond-breaking; xi) Molecular excitations,
ionizations, radiationless transitions, and H-atom migration \cite{chem-eff}
in molecules contaning a nucleus which experiments $\beta $-decay; xii) The
origin of biological chirality has been postulated to be the natural 
$\beta^{-}$-decay of stellar $^{14}C$ and $^{14}N$ and gases built with these
isotopes \cite{Primakoff}; xiii) Energy splitting in molecules due to parity
violation effects in molecules resulting from the electroweak interaction
serve as a guide for accurate experiments \cite{Berger}; xiv) Mechanisms of 
$\beta$-decay in living systems are used in cancer research and biomedicine,
where radioisotopes are incorporated in molecules and delivered to viruses,
phages or cells \cite{chem-beta}.

The weight of the atomic effects in nuclear $\beta$-decay has not been
sufficiently investigated yet. Theoretical results estimate that the shape of
the spectra, life-times and rates of decay in atoms are small but still
observable \cite{Chizma}. Atomic effects were studied theoretically with the
pioneering work of Feinberg \cite{Feinberg}, Migdal \cite{Migdal1,Migdal2} and 
Levinger \cite{Levinger}, 
and experimentally \cite{Freedman}. The atomic effects in many-electron
atoms during nuclear $\beta $-decay has been investigated by several authors 
\cite{Pyper}, including the $\beta$-induced excitation and ionization final
atom one-electron atoms \cite{Schwartz}. Skorobogatov \cite{Skorobogatov}
calculated the probabilities of ionization for the atoms from Li to Kr using
hydrogenlike Slater orbitals and Hartree-Fock (HF) wave functions. Recently, 
Frolov and Talman \cite{FT,TF} calculated the transition probabilities from
ground to ground state of the atoms He to Ar using the relativistic
HF method. Note that HF wave functions can be used only for the
calculation of transition probabilities of atoms and molecules from ground to ground state. 
Theoretical investigations introducing electronic correlation are reported in Refs. 
\cite{Winther,Kolos-chem,Skorobogatov,Mukoyama1,Mukoyama2}. In this work we
calculate the atomic effects with the highest possible accuracy using
correlated quantum mechanical methods.

\subsection{Validity of the Sudden Approximation}

In general, the initial atom is not a bare nucleus. The atom can be in its
ground or any of its excited states. The final atom/ion can be also possibly found
in its ground or any excited state. The meaning of the 'sudden
approximation' is the following: the $\beta$-decay process occurs in a much
faster time than the periods of the orbital electrons, or what it is the same, 
the $\beta$-electron leaves the nucleus with quasirelativistic speed, much
faster than the velocity of the orbital electrons. These electrons suffer then
an abrupt (sudden) change of charge from the nucleus and have to rearrange
themselves within the existing orbitals of the final atom/ion or molecule.

During $\beta$-decay the orbital electrons can be ejected by the 
$\beta$-electron in two processes: In the `shake off' process the change of the
nuclear charge in one unity changes the trajectories of the orbital
electrons. In this process most of the orbital electrons jump into new
orbitals of the final atom/ion with $Z+1$. But some number of electrons
cannot jump into new orbitals of the final atom/ion. Then these electrons
are shaked out and the atom is ionized. The second process is the `direct
collision' in which the outgoing $\beta $-electron interacts electromagnetically
with the orbital electrons. Consequently, the electrons are knocked out and
the electron is ionized \cite{Morita}. The theory of ionization was
introduced by Migdal \cite{Migdal1,Migdal2}, Feinberg \cite{Feinberg}, 
\cite{Skorobogatov} and Schwartz \cite{Schwartz} and recently reinvestigated by
Frolov \cite{Frolov,ionization} and Wauters et al. \cite{WV2}.

Let us consider the matrix element $\left| M_{f,i}\right| ^2$ containing the
whole wave functions of the system 'whole-atom-decay transition' \cite{Freedman}:


\begin{multline}
\left\vert M_{f,i}\right\vert ^{2}=\left\vert \int \Psi _{f}^{\ast
}V_{p}\Psi _{i}d\tau \right\vert ^{2} \\
=\left\vert \left\langle \Psi _{f,nuclear}\cdot \Psi _{f,\beta}\cdot 
\Psi _{f,\nu }\cdot \Psi
_{f,atomic}\left\vert V_{p}\right\vert \Psi _{i,nuclear}\cdot \Psi
_{i,atomic}\right\rangle \right\vert ^{2}
\end{multline}
if the $\beta$-electron and neutrino do not collide with the orbital
electrons, and taking into account that the operator does not depend on the
orbital electron position, we can separate the nuclear from the atomic
electron part (similarly to the Born-Oppenheimer approximation, the error
introduced by this approximation is very small \cite{Robertson}) and then the 
integral is separated into two integrals:


\begin{multline}
\left| M_{f,i}\right| ^2=\left| \left\langle \Psi _{f,nuclear}\cdot \Psi
_{f,\beta }\cdot \Psi _{f,\nu }\cdot \Psi
_{f,atomic}\left| V_p\right| \Psi _{i,nuclear}\cdot \Psi
_{i,atomic}\right\rangle \right| ^2 \\
=\left| \left\langle \Psi _{f,nuclear}\cdot \Psi _{f,\beta }\cdot
\Psi _{f,\nu }\left| V_p\right| \Psi _{i,nuclear}\cdot \Psi
_{i,atomic}\right\rangle \right| ^2\cdot \left| \left\langle \Psi
_{f,atomic}|\Psi _{i,atomic}\right\rangle \right| ^2 \\
=\left| M_{f,i}^{^{\prime }}\right| ^2\cdot \left| \left\langle \Psi
_{f,atomic}|\Psi _{i,atomic}\right\rangle \right| ^2
\end{multline}

The overlap integral $\left\langle \Psi _{f,atomic}|\Psi_{i,atomic}\right\rangle$ 
is called 'imperfect overlap' or 'sudden approximation'. As in the photoionization 
the selection rules are $\Delta L=\Delta J=0$. The sudden approximation can be understood 
as a sudden change of the nuclear charge, when the $\beta$-electron leaves the nucleus
and cross through the orbital electrons of the final atom/ion 
instantaneously. The sudden approximation is very accurate (the introduced
errors are $\approx 5\cdot 10^{-4}$ \cite{Watanabe} in the case of tritium
atom, which is the standard system to estimate the accuracy of the sudden
approximation) when the speed of the $\beta$-electron is much higher 
than the speed of the electrons of the orbitals ($v_\beta \gg v_{e(orbitals)}$). 
This is true in decays where the $\beta$-electron is very fast, quasirelativistic, 
with speed close to the $c$, the speed of light, and in light atoms from H to Ne, where 
relativistic effects are small. Note that the sudden approximation neglects the Coulomb interaction
between the $\beta$-electron and the orbital electrons. It also neglects
the 'recoil effects' of the nucleus of the final atom/ions, which are very
small in the process of $\beta$-decay. Recoil effects are important in
nuclear reactions when the fragments are moving \cite{nucl-react,Galilean}.

The overlap integral of Eq. (9) is the amplitude $A_{f,i}$ and its square
the transition probability $P_{f,i}$ \cite{Skorobogatov}:


\begin{equation}
A_{f,i} =\left\langle \Psi _{f,atomic}|\Psi _{i,atomic}\right\rangle ,
\end{equation}


\begin{equation}
P_{f,i} =\left\vert \left\langle \Psi _{f,atomic}^{\ast }|\Psi
_{i,atomic}\right\rangle \right\vert ^{2}.
\end{equation}
the index $f$ (final) is a bounded or continuum state. $P_{0,0}$ is the
ground to ground transition probability. There are some rules in the
calculation of transition probabilities. The probabilities are numbers smaller 
than one, i.e. $0<P_{f,i}<1$. The probability of transition from the initial
ground state to a given excited state is much greater than the one to the
following excited state: $P_{0\rightarrow n}\gg P_{0\rightarrow n+1}$. If
we consider all transition probabilities of the one state of the initial
atom to all states of the final atom, the sum of these probabilities equals
the unity. The sucessive ground and
excited states are the roots of the eigenfunction equation of the Hamiltonian, 
and therefore orthogonal. The sum of all transition probabilities involving the same initial state is one. 
The sum of all probabilities from a given ground or excited
initial states $i$ to all $n$-states of the final atom/ion is the total 
transition probability $P_{\rm total}$: 


\begin{equation}
P_{\rm total}=\sum_{f=1}^nP_{f,i}
\end{equation}
and its difference to the unity is defined as the probability of ionization $P_{i,ion}$:


\begin{equation}
P_{i,ion}=1-P_{total}=1-\sum_{f=1}^nP_{f,i}
\end{equation}
The probability of ionization can be measured experimentally \cite{exp}. The
probability of transition and ionization from the K-shell is defined as:


\begin{equation}
P_{K,ion}=1-P_{0,0}
\end{equation}
where $P_{0,0}$ is the transition probability from ground to ground state.

\subsection{ Necessity of wave functions of very good quality}

The difficulty of calculating the transition probabilities during $\beta$-decay processes, 
even neglecting electromagnetical interactions and recoil effects, is the
necessity of sufficiently accurate wave functions for the ground and excited
states of the initial and final atoms \cite{Schwartz}. Therefore the first
such calculations employed highly accurate Hylleraas-type wave functions 
\cite{Skorobogatov,Kolos1,Kolos2}. Kolos found \cite{Kolos-chem} that the values obtained with wave
functions within the one-electron approximation lead to smaller transition
probabilities than the correlated wave functions (Hylleraas-type \cite{Hylleraas}).
Nevertheless a real proof of the assumption of necessity of very accurate wave functions  
cannot be found in the literature. In this work we shall try to demonstrate this 
assumption and determine the accuracy of the wave functions which is sufficient for the description 
of the atomic effects in $\beta$-decay processes.    

With the development of modern quantum chemistry a variety of methods are available for such purpose.
Glushkov \cite{Glushkov} in his review treating the chemical environment
effect in $\beta $-decay parameters pointed out: ''The wide-spread quantum
mechanical methods (such as the Hartree-Fock (HF) method, the random-phase
approximation, the Coulomb approximation (CA), the Hartree-Fock-Slater (HFS)
and Dirac-Fock (DF) methods, DFT, etc.) are usually used in the atomic
calculations and calculations of the $\beta $-allowed (superallowed)
transitions parameters (...). The difficulties of the corresponding
calculations are well known (insufficiently correct account for exchange and
correlation in the wave functions of $\beta $-particle, problem of gauge
invariance, generation of the non-optimized bases of the wave functions for
a discrete spectrum and continuum, etc.). The nuclear, relativistic,
radiative corrections should be accurately taken into account too.''

Therefore there is a real need of very accurate wave functions which account for the
effect of electron correlation to treat the atomic and chemical effects in $\beta$-decay 
and other nuclear reactions. The most accurate and
well-behaving wave functions are not available for all atoms. For atoms up
to three-electrons Hylleraas (Hy) and exponential Hylleraas (E-Hy) wave
functions can be used. For atoms with more than three-electrons the more
reasonable choice is the Hylleraas-Configuration Interaction method (Hy-CI), 
the Configuration Interaction method (CI) using Slater orbitals, Spin
and angular momentum L$^2$ eigenfunctions and the Multiconfigurational Hartree-Fock method (MCHF). 

In the sudden approximation the angular momentum $L$, electron spin $S$ and
spatial parity $\pi$ of the atomic wave function $\Psi$ are conserved
during the nuclear $\beta^{-}$-decay. Therefore, all approximate wave
functions must be constructed as the eigenfunctions of the operators of
angular momentum $\hat{L}^2$ and total electron spin $\hat{S}^2$.

In addition to the necessity of including correlation effects into the wave
function to improve its accuracy, the use of exponential decaying orbitals
like the Slater-type Orbitals (STO) with the appropriate shape at the
nucleus and far of the nucleus should plays an important role. For a
description of the properties of STO orbitals, see Ref. \cite{STO-rev}.

\section{Methods}

\subsection{The CI and Hy-CI wave functions}

In this study we use the variational CI and Hy-CI wave functions \cite{HYCI1,HYCI2}. 
In general, wave functions of Hylleraas-type expansion converge rapidly to the exact wave
functions. The CI and Hy-CI wave functions can be summarized in the
following expression


\begin{equation}
\Psi =\sum_{p=1}^{N}C_{p}\Phi_{p},\qquad \Phi_p=\hat{O}(\hat{L}^2)\hat{\mathcal{A}}
\phi _p\chi 
\end{equation}
The Hy-CI and CI wave functions are linear combinations of $N$ symmetry
adapted configurations $\Phi_p$, the coefficients $C_p$ are determined
variationally, $\hat{\mathcal{A}}$ is the antisymmetrization operator and $\chi$ is a 
spin eigenfunction. The spatial part of the basis functions consists of Hartree products of Slater
orbitals. In the case of the Hy-CI the Hartree products are multiplied by up to one interelectronic coordinate 
$r_{ij}$  


\begin{equation}
\phi _p=r_{ij}^\nu \prod_{k=1}^n\phi _k(r_k,\theta _k,\varphi _k),
\end{equation}
where $\nu =0,1$ are employed for CI and Hy-CI wave functions, respectively.
The basis functions $\phi _p$, are products of $s$, $p$, $d$, $f$, or higher angular momentum 
Slater orbitals. For more details about the construction 
of CI and Hy-CI wave functions, see Ref. \cite{FR3}.  

The difference between the Hy-CI and CI wave functions is that the first contains 
\textit{explicitly} $r_{ij}$ coordinates into the wave function, whereas in the CI
wave function these coordinates are included \textit{implicitly}. The CI
wave function does contain terms $r_{ij}^2$, $r_{ij}^4$, $\cdots$ 
$r_{ij}^{2n}$ which can be formed by combination of angular orbitals $p$, $d$, 
$f$, $\cdots $. Therefore the Hy-CI wave function fulfill the electronic cusp
condition \cite{RW}:


\begin{equation}
\left( \frac 1\Psi \frac{\partial \Psi }{\partial r_{ij}}\right)
_{r_{ij}=0}=\frac 12.
\end{equation}
For contrary the CI wave function does not fulfill this condition. But the
nuclear cusp condition is always fulfill, in the CI, Hy-CI\ as in the
HF wave functions:


\begin{equation}
\left( \frac 1\Psi \frac{\partial \Psi }{\partial r_i}\right) _{r_i=0}=-Z,
\end{equation}
$Z$ is the atomic charge, or the orbital exponent. The cusps (positive for
repulsion and negative for attraction) account for two-body correlation, but
not for three-body correlation.

These conditions are a result of the singularities of the Hamiltonian at $%
r_i=0$ and $r_{ij}=0$, which are i.e. for the case of the He atom:


\begin{equation}
H=-\frac 12\sum_{i=1}^2\frac{\partial ^2}{\partial r_i^2}-\sum_{i=1}^2\frac
1{r_i}\frac \partial {\partial r_i}-\sum_{i=1}^2\frac 2{r_i}+\frac
1{r_{12}}-\frac 2{r_{12}}\frac \partial {\partial r_{12}}-\frac
12\sum_{i\neq j}^2\frac{r_i^2+r_{12}^2-r_j^2}{r_ir_{12}}\frac{\partial ^2}{%
\partial r_i\partial r_{12}}.
\end{equation}
As the exact wave function is obtained from the equation: $H\Psi /\Psi =E$.
This equation leads to the exact energy only if the cusps conditions of Eqs.
(17,18) are fulfilled.

This work is a non-relativistic treatment of the atomic system involved in
the $\beta$-decay. This level of approximation is suitable for light atoms. In the case 
of heavy multicharged ions i.e. bound-state $\beta$-decay in heavy atoms \cite{Taka} relativisitic effects 
should be included to the order of $\alpha^2$, being $\alpha$ the fine structure constant, and 
relativistic wave functions are used like Dirac-Kohn-Sham relativistic wave functions \cite{Glushkov-mu}. 

\subsection{Slater orbitals \& Gaussian orbitals}

Slater-type orbitals \cite{Slater} are considered as the natural basis
functions in quantum molecular calculations. They resemble the true
orbitals, since Slater orbitals (monomials) are a simplification of the
hydrogen-like orbitals (polynomials), which are eigenfunctions of the atomic
one-electron Schr\"{o}dinger equation. The Slater orbitals contain a factor 
$e^{-\alpha \cdot r}$. They are defined


\begin{equation}
\phi (\mathbf{r})=r^{n-1}e^{-\alpha r}Y_l^m(\theta ,\varphi ).
\end{equation}
the parameter $\alpha$ is the adjustable variable (for each orbital)
and $Y_l^m(\theta ,\varphi )$ are the complex spherical harmonics. With this
functional form the Slater orbitals represent well the electron density near the
nucleus (cusp) and far from the nucleus (correct asymptotic decay).
Conversely, the Gaussian orbitals (exponential form $e^{-\alpha \cdot r^2}$)
have erroneous shape near and far from the nucleus (no cusp). Also far of
the nucleus the Gaussian orbitals tend to zero much faster than Slater ones.
Finally, to reproduce a single Slater orbital many Gaussian orbitals are
necessary, but the electron cusp at the nucleus is still missing.
For more information, see Ref. \cite{STO-rev}

$\beta $-decay is a nuclear process. We are interested in the redistribution
of the electronic density during $\beta $-decay. For a correct description
of the electronic effects near of the nucleus (electron-cusp) we need
orbitals which behave well near of the nucleus, such as Slater orbitals.

\subsection{The $\beta $-decay in the He atom}

The $\beta$-decay of the $^6$He atom is a good example to illustrate the
role of the accuracy of the wave functions (determined by different methods)
in the calculation of the atomics effects. The $\beta$-decay of He atom 


\be 
He\rightarrow Li^{+}+e^{-}+\overline{\nu _e}.
\ee 

In a previous paper \cite{FR1} we have calculated the transition probabilities 
of the transitions from the $^1$S ground state of He atom to the first ten low-lying $^1$S ground 
and excited states of the Li$^{+}$ ion using the Hy-CI method, and with
them the probability of ionization to the Li$^{2+}$ ion. The values of these
probabilities have been also determined by the CI method using B-splines
orbitals by Wauters and Vaeck \cite{WV} and by the full relativistic
four-component Dirac-Kohn-Sham (DKS) method by Glushkov \cite{Glushkov}.
Note that B-splines are basis set constructed with polynomials which
represent well the true shape of the orbitals. Also the Slater orbitals used
in the Hy-CI calculations are the natural (best) orbitals for quantum
mechanical calculations.

The energies of the involved states obtained by the three different methods
can be compared together with the corresponding transition probabilities,
see Table I. Note that the energy is one of the properties obtained from
wave functions, and therefore one can take the accuracy of the energy values as a measure of the
accuracy of the wave functions. For the ground state of He atom (accuracy 
$\approx 1.0\cdot 10^{-9}$ a.u.) and the lowest low-lying states of Li$^{+}$
(accuracy $\approx 1.0\cdot 10^{-7}$ a.u. ) the Hy-CI values are the most
accurate. Hylleraas-type calculations use to serve as reference for other
methods. For higher excited states our calculation method is not as accurate as 
for the lower excited states. The reason is that we have used one optimized set 
of exponents, and for higher excited states more flexibility in the selection of the orbital 
exponents would be necessary. The details of our Hy-CI calculations on the He atom 
and Li$^+$ ion can be found in Ref. \cite{FR1}.

The transition probabilities were determined using Eq. (11). The transition
probability of the transition from the ground state of He atom to the ground state of Li$^{+}$
calculated by the Hy-CI method is $70.86$ $\%$. This value is very close to $70.85\%$
obtained by the CI (B-splines) method, to $70.84\%$ by the Multiconfigurational
Hartree-Fock method (MCHF) \cite{Froese}, and to the value $68.13\%$ by the DKS method.
The transition probabilities of transitions from the ground state of He atom to the first
and second excited states of the Li$^{+}$ atom using the Hy-CI method are
14.94 \% and 1.86 \%, respectively. These values agree very well with the ones by Wauters
and Vaeck and Glushkov. Higher transition probabilities decrease rapidly,
but their calculation is important for the calculation of the total transition
probability Eq. (12) and the probability of ionization Eq. (13). The 
comparison of the calculations using several methods are summarized in Table
II.

The values of the total probability of transition are $89.21\%$ by the Hy-CI
method, $89.09\%$ by the CI method with B-splines and $87.04\%$ by the DKS
method. All three values agree well with the experimental value of Carlson 
$89.9\pm 0.2\%$. The probability of ionization using Hy-CI wave function is 
$10.79\%$, $7.47\%$ using CI method with B-splines and $9.85\%$ by the DKS 
method. The results are very accurate compared with the experimental value
of the single ionization $10.40\pm 0.2\%$ determined by Carlson \cite{exp}
(radioactive recoil spectrometry).

For the probability of ionization of an electron from the K-shell, Eq. (14),
we have obtained a $29.14\%$ by the Hy-CI method compared with $28.99\%$
from Skorogobatov using the Hylleraas method, $29.15\%$ by the B-splines
(CI) method \cite{WV}. Note that the excitation of an electron of the
K-shell creates a vacancy.

In the calculation of an additional ionization after $\beta $-decay there is
less agreement. The calculation of the ionization of a further electron
during $\beta$-decay is very complex. Some steps in this direction are
discussed in Ref. \cite{ionization}. Further in Table II some values of
transition probabilities calculated by several authors are given for
completeness. Winther \cite{Winther} employed wave functions which were not
orthogonal. In the second part of Table II transition probabilities using
single-configuration (SCF) wave functions using different kinds of orbitals
are presented. The values of the obtained transition probabilities are
smaller than in the case of using methods which account for electron
correlation.

Three facts explain the complete agreement of the calculations of the atomic
effects in the $\beta$-decay of the He atom by the mentioned correlated
quantum mechanical methods: (1) the agreement accuracy of the energies of at
least $\approx 1.0\cdot 10^{-4}$ a.u. is sufficient to ensure an accuracy of 
$0.01\%$ in the values of the transition probabilities. The accuracy in the
total energy is a consequence of the inclusion of electron correlation (in
different ways) into the wave functions; (2) the good results are due to the
use of well-behaving orbitals as basis functions. In the case of the Hy-CI
method well-behaving Slater orbitals are used; in the case of B-splines,
these orbitals are linear combinations of polynomials which resemble the
true orbitals. (3) $\beta $-decay is a nuclear process, where the change of
the nuclear charge plays the most important role in the account of the
atomic effects. In nuclear $\beta $-decay is therefore decisive the correct
description of the electron density near of the nucleus.

\section{The transition probabilities in the $\beta$-decay of 
the Li atom calculated by the CI\ and Hy-CI methods}

In a previous paper \cite{FR2} we have also calculated the transition
probabilities during the $\beta$-decay in the Li atom:


\be 
Li\rightarrow Be^{+}+e^{-}+\overline{\nu _e}
\ee
using accurate Hy-CI wave functions for several bound states of the Li atom
and Be$^{+}$ ion. The initial and final states were ground or excited states
of S-, P-, and D-symmetry. In addition, using the CI method we have
determined the low-lying bound S-, P-, D-, F-, G-, H- and I-states of the Li
atom and Be$^{+}$ ion with energies below the corresponding limits of
ionization, which are -7.27991 34126 69305 96491 810(15) a.u. for the Li
atom and -13.65556 62384 23586 70207 810(15) a.u. \cite{Frolov-Li+} for the
Be$^{+}$ ion. The energies of 28 bound states of Li atom and Be$^{+}$ are
listed in Tables III and IV. The details of these energy calculations
employing the CI and Hy-CI methods are given in Ref. \cite{FR3}. Note, that
the additional F-, G-, H- and I-states could have been also calculated by
the Hy-CI method, which is general for any atom, but some kinetic energy
integrals including higher angular orbitals need to be extended in our
actual computer program.

The energy results of Tables III and IV show that the energy differences
between the CI energies and the exact non-relativistic energies are $\approx
1.0-2.0\cdot 10^{-3}$ a.u. (1-2 millihartrees) for all lower and higher
excited states. The Hy-CI energy results are more accurate for the lower
states ( $\approx 1.0\cdot 10^{-6}$a.u. for ground states and $\approx
1.0\cdot 10^{-5}-10^{-4}$ a.u. for lower excited states). For higher excited
states the accuracy slightly decreases to $\approx 1.0\cdot 10^{-3}$ a.u.
due to a restriction in our calculation technique \cite{FR3}. Therefore the
agreement of the energies by different methods is an overall of $\approx
1.0\cdot 10^{-3}$ a.u. for all states of Li and Be$^{+}$ atom. In this work
we calculate the transition probabilities during the $\beta$-decay of the
Li atom using CI wave functions and compare the values with
the available values from the Hy-CI method presented in Ref. \cite{FR2}. The
results will be presented in the next Section. First, let us discuss some
computational aspects of the calculation of transition probabilities.

In the calculation of the overlap integral, also called 'amplitude' of the transition, we
need to evaluate all single overlap integrals between Slater determinants
and multiply appropriatelly by the coefficients of the wave functions. This is
done by calculating the overlap matrix between two states, and summing up
all elements of the overlap matrix multiplied by the corresponding wave
function coefficients.

Recently we have improved our earlier method of calculation of the final
state probabilities during the nuclear $\beta $-decay employed in Ref. \cite
{FR1}. Now, we calculate the overlap between the wave functions of different
length \cite{FR2}. This overlap is the sum of the matrix elements of a
rectangular overlap matrix, multiplied by the corresponding coefficients.
This method of calculation has several advantages, (1) there is not 
the restriction the wave functions of all states should be constructed
using the same basis of orbitals or the same configurations, both may differ
from state to state (2) there are several possible checks for the correctness
of the calculations, like the permutation symmetry of the overlap matrix 
$\langle \Psi _1|\Psi _2\rangle =\langle \Psi _2|\Psi _1\rangle $ and its
unit-norm condition, i.e. $\langle \Psi _1|\Psi _1\rangle =\langle \Psi
_2|\Psi _2\rangle = 1 $. Note that overlaps between states of different spatial
symmetry or different spin states are zero, what has been also tested in
actual calculations.

We have written a computer program in Fortran 90 for the calculation of the
transition probabilities during $\beta$-decay. In the case of using CI wave
functions, it is sufficient the use of double precision arithmetic (about 15
decimal digits in our computer are meaninful) and in the case of the Hy-CI method quadruple
precision is necessary (about 30 decimal digits are accurate), since the overlap 
between Hy-CI wave functions involves
some types of three-electron integrals which accurate evaluation requires
quadruple precision. 

In order to calculate the transition probabilities during $\beta$-decay we
have to evaluate Eq. (11) which is the 'overlap integral' between two
states. Previously we need to determine the accurate wave functions of every
state. For this we solve the Schr\"{o}dinger equation for the state under study and determine
the eigenvalue (energy) and eigenvector (expansion coefficients of the configurations).
The eigenvectors within a same eigenvalue problem are orthogonal. 
Here one has to point out that
the wave functions of this work $\Psi _{f,i}$ with $i=0,\ldots , n$ are eigenfunctions of different
eigenvalue equations of the system Be$^{+}$ ion, since we have optimized the orbital exponents specially for every 
excited state. Therefore the
wave functions of these states must not be strictly orthogonal beetween themselves.
In the practice these eigenvectors are also orthogonal                                              
with an error of $\approx 0.5\cdot 10^{-4}$. We have checked the orthogonality of different final 
states of the Be$^+$ ion using both CI and Hy-CI wave functions. Consequently, in this work we do not orthogonalize
additionally the resulting wave functions before calculating the transition probabilities. 
The sum of the transition probabilities from a state of the initial atom to all the states 
of the same symmetry of final atom is also in this case equal the unity. Note that unnecessary 
transformations can lead to loss of accuracy. Our calculations using non-orthogonal wave functions 
have been tested in the case of the He atom, where our results agree with the ones of the literature, as shown in 
the previous Section. 

The results of the comparison of the transition probability distributions (in $\%$) of S, P and D states 
calculated using the CI and Hy-CI methods, see Tables V-VII, show a very good agreement ($\approx 0.15 \%$ average deviation) 
between the values calculated by both methods CI and Hy-CI. For the lowest energy states the agreement is excellent 
$\approx 0.01\%$ while the largest differences ocurr for the highest $8 ^2S$, $8 ^2P$, $8 ^2D$ states where our method 
of calculation (exponent restriction) is not as accurate as for low states. Several probability distributions 
are nearly complete (some high excitations with very small probabilty are missing) and in these cases it has been 
possible to calculate the total probability and the probability of ionization. 
The CI method has been applied then to calculate the 
transition probability distributions of the F, H and I states, see Tables VIII-X.

\section{Conclusions}

In this paper we review the atomic effects induced by the orbital electrons
during whole-atom-nuclear $\beta$-decay and discusse the accuracy of the
wave functions which are needed. Traditionally it has been assumed that for
the correct calculation of the transition probabilities correlated wave
functions are necessary but a real proof of this has never been shown. In this
work we demonstrate that with the inclusion of electron correlation we
obtain an agreement of $0.01\%$ between methods including electron correlation in the
calculated transition probabilities for ground and low-lying states. We have obtained an agreemenet 
of $\approx 0.01\%$ with the experimental probability of ionization. In the
case of He atom several methods including electron correlation lead to the same results. These values are 
larger than the ones obtained by uncorrelated methods. This proofs the necessity of inclusion of electron
correlation into the wave functions. Hylleraas-type methods lead to the most accurate results and may serve 
as a reference or model for larger systems. Among the Hylleraas-type methods, the Hy-CI method can be used 
for light atoms from He to Ne. Nevertheless electron correlation is not the unique 
effect which has to be taken into account, the use of well-behaving orbitals
which represent well the electron density at the nucleus is also decisive,
note that in the best calculations of the literature Slater and B-splines have been used. 
Finally, one has to point out that $\beta$-decay is
a nuclear process, where the change of the nuclear charge plays the most
important role in the account of the atomic effects. In nuclear $\beta $%
-decay is therefore decisive the correct description of the electron density
near of the nucleus. In other types of $\beta$-decay like bound-state $\beta $-decay
and electron capture the correct description of the electron correlation and electron density at the nucleus are 
still more needed. 

In the case of the Li atom we have demonstrated that there are different
distributions of transition probabilities between the S, P, D, ... I 
states and they follow the same pattern. The largest transition probability during 
$\beta^-$-decay is the one from an initial states $n$ to an final state $n+1$. This
can be explained with the hypothesis that after the sudden change of atomic
of nuclear charge the orbitals of the final atom get more contracted (stable), then the orbital 
electrons of the final atom pass to occupy orbitals of lower energy, 
but not all electrons can be accomodated or fill the new orbitals, there are spin and 
spatian restrictions, therefore one or more electrons would stay in a orbital of about the 
same energy than the original orbital. This orbital would become now 
an excited one in the final atom, so that the most probable state of the final atom is always the 
excited state which is an unity higher than the initial state.  
Another interesting fact is that in the probability distributions where it was
possible to calculate the probability of ionization, this takes the value
about $\approx 14\%$ for all kind of states, as we pointed out before in Ref. \cite{FR2}. 

Finally, in the case of the Li atom, we have repeated the calculations of the 
transition probability distributions between states of S, P, and D using CI wave functions
obtaining as a result that an accuracy of $\approx 1.0\cdot 10^{-3}$ in the calculation of the 
energy of the states ensures an accuracy of $0.01\%$ in the values of the probabilities (an average of $0.15\%$).
  
In addition we have carried out new calculations of the transition probabilities for F, G, and H states, 
They follow the same patern than the other S, P, and D distributions. 
As a conclusion, the CI and MCHF methods can be used to calculate the transition probabilities 
of transitions from initial ground and excited states to final ground and excited states of larger atoms and ions. 
As the CI method using Slater orbitals (and equivalent methods like MCHF) are computationally less expensive, 
they can be applied to larger systems which maybe interesting in nuclear medicine. 

\section{Acknowledgments}

I am deeply indebted to Alexei Frolov for awakening my interest on this fascinating field,  
which has resulted in several common publications. Interesting discussions with J\"orn Manz on 
quantum mechanical methods and the physics of $\beta$-decay have stimulated much of this work.  
I would like to thank the anonimous Referee for generous comments which have served to improve this 
manuscript.

\newpage


\begin{table}[tp]
\begin{center}
\caption{Comparison of the transition probabilities during the $\beta^-$-decay from the ground 
state of He atom to the ground and excited states 
of the Li$^+$ ion calculated by the Hy-CI, B-splines (CI) and DKS methods. 
The energy is in a.u. Prob. is the transition probability in $\%$. 
The calculated energy of the ground state of He atom is -2.903 724 376 99 a.u. ($<1\times 10^{-9}$ a.u. accurate) 
by using the Hy-CI method and -2.903 309 69 a.u. by the B-splines.} 
\scalebox{0.85}{%
\begin{tabular}{ccccccc}
\hline\hline
State \quad & \quad E(Hy-CI) \cite{FR1} \quad & \quad Prob.(Hy-CI) \quad  & \quad E(B-splines) \cite{WV} \quad & \quad Prob.(B-splines) \quad & 
\quad E(DKS) \cite{Glushkov}& \quad Prob.(DKS) \\
\hline
1s1s $^1$S & -7.279 913 407 & 70.86 & -7.279 3492 & 70.85 & -7.279 5438 & 68.13 \\
1s2s $^1$S & -5.040 876 744 & 14.94 & -5.040 8201 & 14.94 & -5.040 8413 & 14.37 \\
1s3s $^1$S & -4.733 755 814 &  1.86 & -4.733 7397 &  1.86 & -4.733 7488 &  1.81 \\
1s4s $^1$S & -4.629 783 493 &  0.62 & -4.629 7767 &  0.62 & -4.629 7798 &  0.63 \\
1s5s $^1$S & -4.582 421 933 &  0.29 & -4.582 4240 &  0.29 & -4.582 4256 &  0.26 \\
1s6s $^1$S & -4.556 877 651 &  0.16 & -4.556 9496 &  0.16 & -4.556 9529 &  0.14 \\
1s7s $^1$S & -4.540 876 955 &  0.11 & -4.541 6882 &  0.10 & -4.541 6916 &  0.10 \\
1s8s $^1$S & -4.528 507 401 &  0.19 & -4.531 8274 &       & -4.531 8322 &       \\
1s9s $^1$S & -4.512 574 267 &  0.07 &             &       &             &       \\
1s10s $^1$S& -4.499 228 362 &  0.11 &             &       &             &       \\
2s2s $^1$S &                &       & -1.904 924  &  1.56 & -1.905 2764 &  1.24 \\
2p2p $^1$S &                &       & -1.628 787  &  0.18 & -1.629 3165 &  0.16 \\
2s3s $^1$S &                &       &             &  0.23 &             &  0.21 \\
2s4s $^1$S &                &       &             &       &             &  0.05 \\
\hline\hline
\end{tabular}}
\end{center}
\end{table}

\newpage


\begin{table}[tp]
\begin{center}
\caption{Comparison of the several transition probabilities during $\beta^-$-decay in the He atom calculated by correlated 
and uncorrelated quantum mechanical methods and the experimental ones. P$_{0,0}$ is the transition probability from ground 
to ground state, P$_{\rm total}$ is the sum of all probabilities from the ground state of the He atom to all bound states of the 
Li$^+$ ion, P$_{\rm ion}$ is the probability of ionization of a single electron and P$_{\rm K,ion}$ is the probability of ionisation 
of one electron of the K-shell.} 
\scalebox{0.90}{%
\begin{tabular}{lcccccc}   
\hline\hline
Method        & \quad P$_{\rm 0,0}$ \quad & \quad P$_{\rm total}$ \quad  & \quad  P$_{\rm ion}$ \quad & \quad 
P$_{\rm K,ion}$ \quad & \quad Double ioniz.$^a$ \quad & \quad Ref. \\
\hline 
Hy-CI            &  70.86 & 89.21  &  10.79 & 29.14  &      & \cite{FR1} \\ 
B-splines (CI)   &  70.85 & 89.09  &   7.47 & 29.15  & 0.32 & \cite{WV}  \\
Skorobogatov (Hy)&        &        &        & 28.991 &      & \cite{Skorobogatov} \\ 
GIDKS            &  68.13 & 87.40  &  9.85  & 31.87  & 0.09 & \cite{Glushkov} \\
MCHF             &  70.84 &        &        &        &      & \cite{WV} \\
Mukoyama (Hy)    &        &        &        & 29.32  &      & \cite{Mukoyama1} \\
Kolos   (Hy)     &        &        &        & 31.41  &      & \cite{Kolos-chem} \\
Winther (Hy)     &        & 89.5$^b$ & 10.5 $\pm$0.5 & 33.2 &   & \cite{Winther}  \\
\hline
experimental     &        & 89.9$\pm$0.2 & 10.4$\pm$0.2 &  &0.042$\pm$0.007 & \cite{exp} \\   
\hline
HF               & 73.1   &        &        &        &      & \cite{FT}   \\ 
hydrogenlike     &        &        &        & 21.72  &      & \cite{Mukoyama1} \\ 
screened hydrog. &        &        &        & 27.53  &      & \cite{Mukoyama1} \\
Weiss            &        &        &        & 26.86  &      & \cite{Weiss}  \\ 
SCF              &        &        &        & 26.9   &      & \cite{Carlson} \\ 
\hline\hline
\end{tabular}}
\end{center}
\footnotetext[1]{Estimated and meassured values of the probability of a second ionisation.}
\footnotetext[2]{Estimated value including extrapolated contributions to bound states.} 
\end{table}

\newpage


\begin{table}[tp]
\begin{center}
\caption{Nonrelativistic energies of the S-, P-, D-, and F-states of the Li atom in a.u. ordered by their stability (n). 
All these states are bound states and lay below the ionisation threshold of the Li$^+$ ion, which is -7.279 913 41 \cite{Frolov-Li+}.   
Details on the Full-CI and Hy-CI calculations and reference energies are given in \cite{FR3}.} 
\scalebox{0.90}{%
\begin{tabular}{cccccl}
\hline\hline
n \quad & \quad Conf. \quad & \quad  State \quad & \quad \quad E(FCI) \quad \quad   & \quad \quad E(Hy-CI) \quad \quad 
& \quad \quad  Ref. E \\
\hline
 1 & 1s$^2$2s & 2$^2S$ & -7.477 192 & -7.478 058 969 & -7.478 060 324 \\
 2 & 1s$^2$2p & 3$^2P$ & -7.408 619 & -7.410 149 407 & -7.410 156 533 \\
 3 & 1s$^2$3s & 3$^2S$ & -7.353 249 & -7.354 093 706 & -7.354 098 421 \\
 4 & 1s$^2$3p & 4$^2P$ & -7.335 658 & -7.337 113 114 & -7.337 151 708 \\
 5 & 1s$^2$3d & 4$^2D$ & -7.334 100 & -7.335 512 623 & -7.335 523 544 \\
 6 & 1s$^2$4s & 4$^2S$ & -7.317 679 & -7.318 517 759 & -7.318 530 846 \\
 7 & 1s$^2$4p & 5$^2P$ & -7.310 383 & -7.311 811 529 & -7.311 889 059 \\
 8 & 1s$^2$4d & 5$^2D$ & -7.309 761 & -7.311 211 047 & -7.311 189 578 \\
 9 & 1s$^2$4f & 5$^2F$ & -7.309 517 &  &  \\
10 & 1s$^2$5s & 5$^2S$ & -7.302 682 & -7.303 496 699 & -7.303 551 579 \\
11 & 1s$^2$5g & 6$^2G$ & -7.299 430 &  &  \\
12 & 1s$^2$2s & 6$^2F$ & -7.299 340 &  &  \\
13 & 1s$^2$5p & 6$^2P$ & -7.298 802 & -7.300 137 068 & -7.300 288 165 \\
14 & 1s$^2$5d & 6$^2D$ & -7.298 502 & -7.299 779 537 & -7.299 927 556 \\
15 & 1s$^2$6s & 6$^2S$ & -7.294 935 & -7.295 739 603 & -7.295 859 510 \\
16 & 1s$^2$6h & 7$^2H$ & -7.293 320 &  &  \\
17 & 1s$^2$6g & 7$^2G$ & -7.293 294 &  &  \\
18 & 1s$^2$6f & 7$^2F$ & -7.293 211 &  &  \\
19 & 1s$^2$6p & 7$^2P$ & -7.292 545 & -7.293 967 122 & -7.294 020 053 \\
20 & 1s$^2$6d & 7$^2D$ & -7.292 387 & -7.293 697 654 & -7.293 810 714 \\
21 & 1s$^2$7i & 8$^2I$ & -7.289 638 &  &  \\
22 & 1s$^2$7h & 8$^2H$ & -7.289 625 &  &  \\
23 & 1s$^2$7g & 8$^2G$ & -7.289 605 &  &  \\
24 & 1s$^2$7s & 7$^2S$ & -7.289 596 & -7.290 231 582 & -7.291 392 237 \\
25 & 1s$^2$7f & 8$^2F$ & -7.289 401 &  &  \\
26 & 1s$^2$7p & 8$^2P$ & -7.288 749 &  &  \\
27 & 1s$^2$7d & 8$^2D$ & -7.288 701 & -7.289 731 555 & -7.290 122 856 \\
28 & 1s$^2$8s & 8$^2S$ & -7.285 695 & -7.286 739 123 &   \\
\hline\hline
\end{tabular}}
\end{center}
\end{table}

\newpage


\begin{table}[tp]
\begin{center}
\caption{Non-relativistic energies of the S-, P-, and D-states of the Be$^+$ ion in a.u. ordered by their stability.
All these states are bound states and ly below the ionisation threshold of the Be$^{2+}$, 
which is -13.655 566 24 a.u. \cite{Frolov-Li+}.
Details on the Full-CI and Hy-CI calculations and reference energies are given in \cite{FR3}.}
\scalebox{0.90}{%
\begin{tabular}{cccccl}
\hline\hline
n \quad & \quad  Conf. \quad & \quad State \quad & \quad \quad E(FCI) \quad \quad    
& \quad \quad E(Hy-CI) \quad \quad  & \quad \quad \quad Ref. E \\
\hline
 1 & 1s$^2$2s & 2$^2S$ & -14.323 769 & -14.324 761 678 & -14.324 763 177 \\
 2 & 1s$^2$2p & 3$^2P$ & -14.177 409 & -14.179 327 999 & -14.179 333 293 \\
 3 & 1s$^2$3s & 3$^2S$ & -13.921 830 & -13.922 784 968 & -13.922 789 269 \\
 4 & 1s$^2$3p & 4$^2P$ & -13.883 425 & -13.885 115 345 & -13.885 15      \\
 5 & 1s$^2$3d & 4$^2D$ & -13.876 447 & -13.878 041 021 & -13.877 871 0   \\
 6 & 1s$^2$4s & 4$^2S$ & -13.797 754 & -13.798 706 849 & -13.798 716 609 \\
 7 & 1s$^2$4p & 5$^2P$ & -13.781 975 & -13.783 570 878 & -13.783 518 3   \\
 8 & 1s$^2$4f & 5$^2F$ & -13.779 946 &  &   \\
 9 & 1s$^2$4d & 5$^2D$ & -13.779 084 & -13.780 558 927 & -13.780 514 4   \\
10 & 1s$^2$5s & 5$^2S$ & -13.743 655 & -13.744 580 355 & -13.744 631 82  \\
11 & 1s$^2$5p & 6$^2P$ & -13.735 466 & -13.736 438 672 & -13.737 18      \\
12 & 1s$^2$5g & 6$^2G$ & -13.735 021 &  &  \\
13 & 1s$^2$5f & 6$^2F$ & -13.734 924 &  &  \\
14 & 1s$^2$5d & 6$^2D$ & -13.734 024 & -13.735 485 794 & -13.735 455 4   \\
15 & 1s$^2$6s & 6$^2S$ & -13.715 222 & -13.716 152 058 & -13.716 286 24  \\
16 & 1s$^2$6h & 7$^2H$ & -13.710 578 &  &  \\
17 & 1s$^2$6g & 7$^2G$ & -13.710 575 &  &  \\
18 & 1s$^2$6f & 7$^2F$ & -13.710 457 &  &  \\
19 & 1s$^2$6p & 7$^2P$ & -13.710 140 & -13.711 935 268 & -13.712 06      \\
20 & 1s$^2$6d & 7$^2D$ & -13.709 538 & -13.709 859 822 &    \\
21 & 1s$^2$7s & 7$^2S$ & -13.697 421 & -13.699 131 127 &    \\
22 & 1s$^2$7i & 8$^2I$ & -13.695 844 &  &  \\
23 & 1s$^2$7h & 8$^2H$ & -13.695 828 &  &  \\
24 & 1s$^2$7g & 8$^2G$ & -13.695 806 &  &  \\
25 & 1s$^2$7f & 8$^2F$ & -13.695 579 &  &  \\
26 & 1s$^2$7p & 8$^2P$ & -13.695 228 & -13.695 922 402 &   \\ 
27 & 1s$^2$7d & 8$^2D$ & -13.694 804 &  &  \\
28 & 1s$^2$8s & 8$^2S$ & -13.684 764 & -13.687 372 394 &    \\
\hline\hline
\end{tabular}}
\end{center}
\end{table}

\newpage


\begin{table}[tp]
\caption{Transition probabilities between states of S-symmetry for the nuclear $\beta^-$-decay
of the Li$^a$ atom to the Be$^+$ ion$^b$. Probabilities from an initial state $i$ to a final state $f$ 
P$_{f,i}$ in $\%$. Diff. are the probability differences in $\%$. }
\begin{center}
\scalebox{0.90}{%
\begin{tabular}{cccrrr}
\hline\hline
States Li $\rightarrow$ Be$^+$ &\quad  Amplitude (Hy-CI) \quad &  Amplitude (CI) \qquad & P$_{f,i}$ (Hy-CI) & \quad P$_{f,i}$ (CI) 
& \quad Diff. \\
\hline
2$^2$S $\rightarrow$ 2$^2$S & 0.759 683 487 & 0.759 397 445 & 57.71 & 57.67 &  0.04 \\
2$^2$S $\rightarrow$ 3$^2$S & 0.514 929 058 & 0.514 902 321 & 26.52 & 26.51 &  0.01 \\
2$^2$S $\rightarrow$ 4$^2$S & 0.073 790 160 & 0.074 804 998 &  0.54 &  0.56 & -0.02 \\ 
2$^2$S $\rightarrow$ 5$^2$S & 0.043 113 179 & 0.042 067 573 &  0.19 &  0.18 &  0.01 \\
2$^2$S $\rightarrow$ 6$^2$S & 0.029 411 301 & 0.028 691 213 &  0.09 &  0.08 &  0.01 \\                        
2$^2$S $\rightarrow$ 7$^2$S & 0.021 688 396 & 0.022 204 212 &  0.05 &  0.05 &  0.00 \\ 
2$^2$S $\rightarrow$ 8$^2$S & 0.017 296 174 & 0.024 954 695 &       &  0.06 &  0.09 \\
P$_{\rm total}$             &               &               & 85.25 & 85.11 &  0.14 \\
P$_{\rm ion}$               &               &               & 14.75 & 14.89 & -0.14 \\
\hline
3$^2$S $\rightarrow$ 2$^2$S & 0.239 962 786 & 0.240 656 991 &  5.76 &  5.79 & -0.03 \\
3$^2$S $\rightarrow$ 3$^2$S & 0.466 529 800 & 0.465 113 426 & 21.76 & 21.63 &  0.15 \\
3$^2$S $\rightarrow$ 4$^2$S & 0.757 456 066 & 0.757 375 068 & 57.37 & 57.36 &  0.01 \\
3$^2$S $\rightarrow$ 5$^2$S & 0.055 586 071 & 0.054 475 234 &  0.31 &  0.30 &  0.01 \\
3$^2$S $\rightarrow$ 6$^2$S & 0.012 740 357 & 0.011 091 305 &  0.02 &  0.01 &  0.01 \\                      
3$^2$S $\rightarrow$ 7$^2$S & 0.013 723 711 & 0.014 795 603 &  0.02 &  0.02 &  0.00 \\ 
3$^2$S $\rightarrow$ 8$^2$S & 0.012 543 893 & 0.021 513 718 &       &  0.05 &  0.20 \\
P$_{\rm total}$             &               &               & 85.49 & 85.16 &  0.33 \\
P$_{\rm ion}$               &               &               & 14.51 & 14.84 & -0.33 \\
\hline
4$^2$S $\rightarrow$ 2$^2$S & 0.132 669 559 & 0.133 270 329 &  1.76 &  1.78 & -0.02 \\
4$^2$S $\rightarrow$ 3$^2$S & 0.236 587 524 & 0.237 980 180 &  5.60 &  5.66 & -0.06 \\
4$^2$S $\rightarrow$ 4$^2$S & 0.122 373 066 & 0.120 054 777 &  1.50 &  1.44 &  0.06 \\
4$^2$S $\rightarrow$ 5$^2$S & 0.828 124 464 & 0.828 363 202 & 68.58 & 68.62 & -0.04 \\
4$^2$S $\rightarrow$ 6$^2$S & 0.277 774 076 & 0.275 419 404 &  7.72 &  7.59 &  0.13 \\                                   
4$^2$S $\rightarrow$ 7$^2$S & 0.007 347 388 & 0.002 133 673 &  0.01 &  0.00 &  0.01 \\ 
4$^2$S $\rightarrow$ 8$^2$S & 0.007 165 497 & 0.011 425 938 &       &  0.01 &       \\
P$_{\rm total}$             &               &               & 85.35 & 85.10 &  0.25 \\
P$_{\rm ion}$               &               &               & 14.65 & 14.90 & -0.25 \\
\hline\hline
\end{tabular}}
\end{center}
\end{table}

\newpage 

\begin{table}[tp]
{Continuation TABLE V.} 
\begin{center}
\scalebox{0.90}{%
\begin{tabular}{cccrrr}
\hline\hline
States Li $\rightarrow$ Be$^+$ &\quad  Amplitude (Hy-CI) \quad &  Amplitude (CI) \qquad & P$_{f,i}$ (Hy-CI) & \quad P$_{f,i}$ (CI)
& \quad Diff. \\
\hline
5$^2$S $\rightarrow$ 2$^2$S & 0.087 318 854 & 0.088 471 255 &  0.76 &  0.78 & -0.02 \\
5$^2$S $\rightarrow$ 3$^2$S & 0.148 984 137 & 0.151 829 221 &  2.22 &  2.31 & -0.09 \\
5$^2$S $\rightarrow$ 4$^2$S & 0.109 684 232 & 0.113 333 243 &  1.20 &  1.28 & -0.08 \\
5$^2$S $\rightarrow$ 5$^2$S & 0.175 864 858 & 0.175 564 518 &  3.09 &  3.08 &  0.01 \\
5$^2$S $\rightarrow$ 6$^2$S & 0.698 154 162 & 0.697 757 378 & 48.74 & 48.68 &  0.06 \\
5$^2$S $\rightarrow$ 7$^2$S & 0.503 067 106 & 0.491 433 474 & 25.31 & 24.15 &  1.16 \\
5$^2$S $\rightarrow$ 8$^2$S & 0.060 646 235 & 0.022 368 017 &       &  0.05 &  0.10 \\
P$_{\rm total}$             &               &               & 81.48 & 80.33 &  1.15 \\
P$_{\rm ion}$               &               &               & 18.52 & 19.67 & -1.15 \\
\hline
6$^2$S $\rightarrow$ 2$^2$S & 0.063 750 613 & 0.062 636 932 &  0.41 &  0.39 &  0.02 \\                            
6$^2$S $\rightarrow$ 3$^2$S & 0.104 007 178 & 0.103 112 719 &  1.08 &  1.06 &  0.02 \\                                       
6$^2$S $\rightarrow$ 4$^2$S & 0.079 072 302 & 0.078 185 058 &  0.63 &  0.61 &  0.02 \\                           
6$^2$S $\rightarrow$ 5$^2$S & 0.071 415 619 & 0.073 692 262 &  0.51 &  0.54 & -0.03 \\                            
6$^2$S $\rightarrow$ 6$^2$S & 0.350 972 033 & 0.356 690 530 & 12.32 & 12.72 & -0.40 \\                            
6$^2$S $\rightarrow$ 7$^2$S & 0.430 715 551 & 0.457 444 994 & 18.55 & 20.93 & -2.71 \\
6$^2$S $\rightarrow$ 8$^2$S & 0.682 123 252 & 0.563 002 732 &       & 31.70 &-20.62 \\
\hline 
7$^2$S $\rightarrow$ 2$^2$S & 0.043 057 699 & 0.054 142 084 &  0.19 &  0.29 & -0.10 \\
7$^2$S $\rightarrow$ 3$^2$S & 0.063 693 794 & 0.080 726 270 &  0.41 &  0.65 & -0.24 \\
7$^2$S $\rightarrow$ 4$^2$S & 0.035 450 799 & 0.049 363 620 &  0.13 &  0.24 & -0.11 \\
7$^2$S $\rightarrow$ 5$^2$S & 0.060 157 796 & 0.064 305 520 &  0.36 &  0.41 & -0.05 \\
7$^2$S $\rightarrow$ 6$^2$S & 0.205 613 808 & 0.232 555 042 &  4.23 &  5.41 & -1.18 \\
7$^2$S $\rightarrow$ 7$^2$S & 0.243 887 102 & 0.299 218 319 &  5.95 &  8.95 & -3.00 \\
7$^2$S $\rightarrow$ 8$^2$S & 0.052 389 853 & 0.203 431 504 &       &  4.14 & -4.08 \\
\hline
8$^2$S $\rightarrow$ 2$^2$S & 0.042 650 127 & 0.067 952 950 &  0.18 &  0.46 & -0.28 \\ 
8$^2$S $\rightarrow$ 3$^2$S & 0.064 215 910 & 0.074 866 606 &  0.41 &  0.56 & -0.15 \\
8$^2$S $\rightarrow$ 4$^2$S & 0.046 928 179 & 0.015 445 544 &  0.22 &  0.02 &  0.20 \\
8$^2$S $\rightarrow$ 5$^2$S & 0.029 188 137 & 0.106 052 617 &  0.09 &  1.12 & -1.03 \\
8$^2$S $\rightarrow$ 6$^2$S & 0.168 756 132 & 0.101 787 496 &  2.85 &  1.04 &  1.81 \\
8$^2$S $\rightarrow$ 7$^2$S & 0.255 470 098 & 0.016 935 140 &  6.53 &  0.03 &  6.50 \\
8$^2$S $\rightarrow$ 8$^2$S & 0.135 800 537 & 0.136 201 556 &       &  1.86 &  2.33 \\
\hline\hline
\end{tabular}}
\end{center}
\end{table}

\newpage


\begin{table}
\caption{Transition probabilities between states of P-symmetry for the nuclear $\beta^-$-decay 
of the Li$^a$ atom to the Be$^+$ ion$^b$. A is the amplitude. Probabilities in $\%$.}
\begin{center}
\scalebox{0.90}{%
\begin{tabular}{cccrrr} 
\hline\hline
States Li $\rightarrow$ Be$^+$ &\quad  Amplitude (Hy-CI) \quad &  Amplitude (CI) \qquad & P$_{f,i}$ (Hy-CI) & \qquad P$_{f,i}$ (CI) & 
\qquad Diff. \\
\hline
3$^2$P $\rightarrow$ 3$^2$P & 0.697 549 959 & 0.696 729 315 & 48.66 & 48.54 &  0.12 \\
3$^2$P $\rightarrow$ 4$^2$P & 0.603 885 572 & 0.604 534 886 & 36.47 & 36.55 & -0.08 \\
3$^2$P $\rightarrow$ 5$^2$P & 0.003 979 607 & 0.003 201 263 &  0.00 &  0.00 &  0.00 \\
3$^2$P $\rightarrow$ 6$^2$P & 0.020 232 690 & 0.016 040 988 &  0.04 &  0.03 &  0.01 \\         
3$^2$P $\rightarrow$ 7$^2$P & 0.013 143 263 & 0.013 378 237 &  0.02 &  0.02 &  0.00  \\                        
3$^2$P $\rightarrow$ 8$^2$P & 0.010 297 817 & 0.010 701 766 &  0.02 &  0.01 &  0.01  \\              
P$_{\rm total}$             &               &               & 85.21 & 85.15 &  0.06  \\ 
P$_{\rm ion}$               &               &               & 14.79 & 14.85 & -0.06  \\
\hline
4$^2$P $\rightarrow$ 3$^2$P & 0.275 908 160 & 0.276 705 772 &  7.61 &  7.66 & -0.05  \\
4$^2$P $\rightarrow$ 4$^2$P & 0.319 479 925 & 0.315 597 281 & 10.21 &  9.96 &  0.25  \\
4$^2$P $\rightarrow$ 5$^2$P & 0.801 261 129 & 0.800 755 115 & 64.20 & 64.12 &  0.08  \\   
4$^2$P $\rightarrow$ 6$^2$P & 0.166 010 974 & 0.180 340 390 &  2.76 &  3.25 & -0.49  \\                        
4$^2$P $\rightarrow$ 7$^2$P & 0.004 047 006 & 0.007 921 346 &  0.00 &  0.01 & -0.01   \\              
4$^2$P $\rightarrow$ 8$^2$P & 0.005 141 118 & 0.003 505 020 &  0.00 &  0.00 &  0.00   \\                  
P$_{\rm total}$             &               &               & 84.78 & 85.00 & -0.22   \\
P$_{\rm ion}$               &               &               & 15.22 & 15.00 &  0.22    \\
\hline
5$^2$P $\rightarrow$ 3$^2$P & 0.161 045 822 & 0.162 629 425 &  2.59 &  2.64 & -0.05   \\                
5$^2$P $\rightarrow$ 4$^2$P & 0.195 960 248 & 0.201 082 812 &  3.84 &  4.04 & -0.20   \\                               
5$^2$P $\rightarrow$ 5$^2$P & 0.046 100 299 & 0.050 845 824 &  0.21 &  0.26 & -0.05   \\      
5$^2$P $\rightarrow$ 6$^2$P & 0.724 469 360 & 0.769 907 028 & 52.49 & 59.28 &  0.21   \\                    
5$^2$P $\rightarrow$ 7$^2$P & 0.425 779 325 & 0.430 229 662 & 18.13 & 18.51 & -0.38   \\                      
5$^2$P $\rightarrow$ 8$^2$P & 0.032 279 556 & 0.015 658 028 &  0.10 &  0.02 &  0.08   \\                        
P$_{\rm total}$             &               &               & 77.36 & 84.75 & -7.39   \\     
P$_{\rm ion}$               &               &               & 22.64 & 15.25 &  7.39   \\     
\hline\hline
\end{tabular}}
\end{center}
\end{table}

\newpage

\begin{table}
{Continuation TABLE VI}
\begin{center}
\scalebox{0.90}{%
\begin{tabular}{cccrrr}
\hline\hline
States Li $\rightarrow$ Be$^+$ &\quad  Amplitude (Hy-CI) \quad &  Amplitude (CI) \qquad & P$_{f,i}$ (Hy-CI) & \qquad P$_{f,i}$ (CI) &
\qquad Diff. \\
\hline
6$^2$P $\rightarrow$ 3$^2$P & 0.113 441 928 & 0.111 196 347 &  1.29 &  1.24 &  0.05   \\                   
6$^2$P $\rightarrow$ 4$^2$P & 0.135 765 197 & 0.138 651 878 &  1.84 &  1.92 & -0.08   \\                          
6$^2$P $\rightarrow$ 5$^2$P & 0.017 086 488 & 0.022 989 302 &  0.03 &  0.05 & -0.02   \\                            
6$^2$P $\rightarrow$ 6$^2$P & 0.328 135 052 & 0.287 332 278 & 10.77 &  8.26 &  2.51   \\                
6$^2$P $\rightarrow$ 7$^2$P & 0.547 505 865 & 0.541 735 488 & 29.98 & 29.35 &  0.63    \\                        
6$^2$P $\rightarrow$ 8$^2$P & 0.638 196 904 & 0.643 629 613 & 40.73 & 41.43 & -0.70    \\                             
P$_{\rm total}$             &               &               & 84.64 & 82.25 &  2.39    \\  
P$_{\rm ion}$               &               &               & 15.36 & 17.75 & -2.39    \\
\hline
7$^2$P $\rightarrow$ 3$^2$P & 0.081 224 665 & 0.081 299 364 &  0.66 &  0.66 &  0.00     \\                
7$^2$P $\rightarrow$ 4$^2$P & 0.099 158 755 & 0.099 999 877 &  0.98 &  1.00 & -0.02     \\                         
7$^2$P $\rightarrow$ 5$^2$P & 0.029 284 955 & 0.027 916 959 &  0.09 &  0.08 &  0.01     \\                             
7$^2$P $\rightarrow$ 6$^2$P & 0.177 208 892 & 0.152 708 579 &  3.14 &  2.33 &  0.81     \\                  
7$^2$P $\rightarrow$ 7$^2$P & 0.353 201 418 & 0.348 418 296 & 12.48 & 12.14 &  0.34     \\                  
7$^2$P $\rightarrow$ 8$^2$P & 0.234 646 011 & 0.218 815 092 &  5.51 &  4.79 &  0.72     \\                                
\hline
8$^2$P $\rightarrow$ 3$^2$P & 0.121 288 399 & 0.063 651 197 & 1.47  & 0.41  & 1.06  \\
8$^2$P $\rightarrow$ 4$^2$P & 0.156 522 176 & 0.077 771 665 & 2.45  & 0.60  & 1.85  \\
8$^2$P $\rightarrow$ 5$^2$P & 0.084 825 989 & 0.026 525 030 & 0.72  & 0.07  & 0.65  \\
8$^2$P $\rightarrow$ 6$^2$P & 0.143 171 882 & 0.096 304 503 & 2.05  & 0.93  & 1.12  \\
8$^2$P $\rightarrow$ 7$^2$P & 0.348 554 021 & 0.236 260 876 & 12.15 & 5.58  & 6.57   \\
8$^2$P $\rightarrow$ 8$^2$P & 0.430 425 119 & 0.240 332 517 & 18.53 & 5.78  & 12.75  \\
\hline\hline
\end{tabular}}
\end{center}
\end{table}

\newpage


\begin{table}
\caption{Transition probabilities between states of D-symmetry for the nuclear $\beta^-$-decay of the Li atom$^a$ to the 
Be$^+$ ion$^b$. A is the amplitude. Probabilities in $\%$.}
\begin{center}
\scalebox{0.90}{%
\begin{tabular}{cccrrr}
\hline\hline
States Li $\rightarrow$ Be$^+$ &\quad  Amplitude (Hy-CI) \quad &  Amplitude (CI) \qquad & P$_{f,i}$ (Hy-CI) & \qquad P$_{f,i}$ (CI) 
& \qquad Diff. \\
\hline
4$^2$D $\rightarrow$ 4$^2$D & 0.613 644 354 & 0.612 485 193 & 37.66 & 37.51 &  0.17 \\      
4$^2$D $\rightarrow$ 5$^2$D & 0.679 777 640 & 0.679 882 548 & 46.22 & 46.22 & -0.60 \\       
4$^2$D $\rightarrow$ 6$^2$D & 0.130 774 298 & 0.131 313 898 &  1.71 &  1.72 & -0.17 \\    
4$^2$D $\rightarrow$ 7$^2$D & 0.005 331 066 & 0.001 828 482 &  0.00 &  0.00 &  0.00 \\ 
4$^2$D $\rightarrow$ 8$^2$D & 0.008 233 705 & 0.002 373 197 &  0.01 &  0.00 &  0.01 \\
P$_{\rm total}$             &               &               & 85.60 & 84.45 &  1.15 \\     
P$_{\rm ion}$               &               &               & 14.40 & 15.55 & -1.15 \\   
\hline
5$^2$D $\rightarrow$ 4$^2$D & 0.297 196 812 & 0.317 734 009 &  8.83 & 10.10 & -1.27 \\        
5$^2$D $\rightarrow$ 5$^2$D & 0.105 347 730 & 0.134 422 801 &  1.11 &  1.81 & -0.07 \\     
5$^2$D $\rightarrow$ 6$^2$D & 0.764 157 229 & 0.772 857 353 & 58.39 & 59.73 & -1.34 \\             
5$^2$D $\rightarrow$ 7$^2$D & 0.314 342 441 & 0.369 966 072 &  9.88 & 13.69 & -3.81 \\
5$^2$D $\rightarrow$ 8$^2$D & 0.002 237 872 & 0.025 890 364 &  0.00 &  0.07 & -0.07 \\
P$_{\rm total}$             &               &               & 78.21 & 85.40 & -7.19 \\
P$_{\rm ion}$               &               &               & 21.79 & 14.60 &  7.19 \\
\hline
6$^2$D $\rightarrow$ 4$^2$D & 0.220 556 194 & 0.205 308 574 &  4.86 &  4.22 &  0.67  \\
6$^2$D $\rightarrow$ 5$^2$D & 0.137 890 137 & 0.144 158 665 &  1.90 &  2.08 & -0.35  \\
6$^2$D $\rightarrow$ 6$^2$D & 0.303 521 357 & 0.206 106 682 &  9.21 &  4.25 &  4.96  \\
6$^2$D $\rightarrow$ 7$^2$D & 0.662 141 399 & 0.612 921 438 & 43.84 & 37.57 &  6.27  \\
6$^2$D $\rightarrow$ 8$^2$D & 0.408 116 762 & 0.598 166 105 & 16.66 & 35.78 & -19.12 \\
\hline\hline
\end{tabular}}
\end{center}
\end{table}

\newpage

\begin{table}
{Continuation TABLE VII.}
\begin{center}
\scalebox{0.90}{%
\begin{tabular}{cccrrr}
\hline\hline
States Li $\rightarrow$ Be$^+$ &\quad  Amplitude (Hy-CI) \quad &  Amplitude (CI) \qquad & P$_{f,i}$ (Hy-CI) & \qquad P$_{f,i}$ (CI)
& \qquad Diff. \\
\hline
7$^2$D $\rightarrow$ 4$^2$D & 0.219 012 937 & 0.147 883 167 &  4.80 &  2.19 & 2.61   \\
7$^2$D $\rightarrow$ 5$^2$D & 0.109 576 684 & 0.118 686 190 &  1.20 &  1.41 & -0.21  \\ 
7$^2$D $\rightarrow$ 6$^2$D & 0.275 765 182 & 0.073 213 641 &  7.60 &  0.54 &  7.06  \\  
7$^2$D $\rightarrow$ 7$^2$D & 0.172 623 479 & 0.343 261 842 &  2.98 & 11.78 & -8.80  \\
7$^2$D $\rightarrow$ 8$^2$D & 0.238 431 045 & 0.298 102 929 &  5.68 &  8.89 & -3.21  \\
P$_{\rm total}$             &               &               & 44.35 & 24.81 & 19.54  \\
P$_{\rm ion}$               &               &               & 55.65 & 75.19 &-19.54  \\
\hline 
8$^2$D $\rightarrow$ 4$^2$D & 0.250 505 561 & 0.114 374 659 &  6.28 &  1.31 & 4.97   \\
8$^2$D $\rightarrow$ 5$^2$D & 0.224 414 920 & 0.097 554 411 &  5.04 &  0.95 & 4.09   \\
8$^2$D $\rightarrow$ 6$^2$D & 0.113 445 739 & 0.028 525 194 &  1.29 &  0.08 & 1.21   \\
8$^2$D $\rightarrow$ 7$^2$D & 0.227 483 595 & 0.213 607 790 &  5.17 &  4.56 & 0.61   \\
8$^2$D $\rightarrow$ 8$^2$D & 0.476 006 811 & 0.286 688 136 & 22.66 &  8.22 &14.44   \\
\hline\hline
\end{tabular}}
\end{center}
\end{table}

\newpage


\begin{table}
\caption{Transition probabilities between states of F-symmetry for the nuclear $\beta^-$-decay 
of the Li atom$^a$ to the Be$^+$ ion$^b$. A is the amplitude. Probabilities in $\%$.}
\begin{center}
\scalebox{0.90}{%
\begin{tabular}{ccr}
\hline\hline
States Li $\rightarrow$ Be$^+$ &   Amplitude (CI) \quad & \quad P$_{f,i}$ (CI)   \\
\hline
\hline
5$^2$F $\rightarrow$ 5$^2$F & 0.543 791 535 & 29.57 \\
5$^2$F $\rightarrow$ 6$^2$F & 0.703 277 972 & 49.46 \\
5$^2$F $\rightarrow$ 7$^2$F & 0.252 544 101 &  6.38 \\
5$^2$F $\rightarrow$ 8$^2$F & 0.013 821 189 &  1.9  \\
P$_{\rm total}$             &               & 87.31 \\
P$_{\rm ion}$               &               & 12.69 \\
\hline
6$^2$F $\rightarrow$ 5$^2$F & 0.335 876 229 & 11.28 \\
6$^2$F $\rightarrow$ 6$^2$F & 0.001 016 736 &  0.00 \\
6$^2$F $\rightarrow$ 7$^2$F & 0.688 168 570 & 47.36 \\
6$^2$F $\rightarrow$ 8$^2$F & 0.503 622 278 & 25.36 \\
\hline
7$^2$F $\rightarrow$ 5$^2$F & 0.232 120 810 &  5.39 \\
7$^2$F $\rightarrow$ 6$^2$F & 0.081 218 845 &  0.66 \\
7$^2$F $\rightarrow$ 7$^2$F & 0.300 585 044 &  9.04 \\
7$^2$F $\rightarrow$ 8$^2$F & 0.434 982 060 & 18.92 \\
\hline
8$^2$F $\rightarrow$ 5$^2$F & 0.177 487 459 &  3.15 \\
8$^2$F $\rightarrow$ 6$^2$F & 0.084 741 247 &  0.72 \\
8$^2$F $\rightarrow$ 7$^2$F & 0.158 306 870 &  2.51 \\
8$^2$F $\rightarrow$ 8$^2$F & 0.333 275 427 & 11.11 \\
\hline
\end{tabular}}
\end{center}
\end{table}

\newpage


\begin{table}
\caption{Transition probabilities between states of G-symmetry for the nuclear $\beta^-$-decay
of the Li atom$^a$ to the Be$^+$ ion$^b$. A is the amplitude. Probabilities in $\%$.}
\begin{center}
\scalebox{0.90}{%
\begin{tabular}{ccr}
\hline\hline
States Li $\rightarrow$ Be$^+$ &   Amplitude (CI) \quad & \quad P$_{f,i}$ (CI)   \\
\hline
\hline
6$^2$G $\rightarrow$ 6$^2$G & 0.483 790 310 & 23.41 \\
6$^2$G $\rightarrow$ 7$^2$G & 0.701 507 008 & 49.21 \\
6$^2$G $\rightarrow$ 8$^2$G & 0.353 179 967 & 12.47 \\
P$_{\rm total}$             &               & 85.09 \\ 
P$_{\rm ion}$               &               & 14.91 \\
\hline
7$^2$G $\rightarrow$ 6$^2$G & 0.345 546 769 & 11.94 \\
7$^2$G $\rightarrow$ 7$^2$G & 0.109 563 505 &  1.20 \\
7$^2$G $\rightarrow$ 8$^2$G & 0.591 706 752 & 35.01 \\
\hline
8$^2$G $\rightarrow$ 6$^2$G & 0.251 079 168 &  6.30 \\
8$^2$G $\rightarrow$ 7$^2$G & 0.018 069 066 &  0.03 \\
8$^2$G $\rightarrow$ 8$^2$G & 0.342 302 378 & 11.72 \\
\hline
\end{tabular}}
\end{center}
\end{table}

\newpage 


\begin{table}
\caption{Transition probabilities between states of H-symmetry and I-symmetry, respectively, 
for the nuclear $\beta^-$-decay
of the Li atom$^a$ to the Be$^+$ ion$^b$. A is the amplitude. Probabilities in $\%$.}
\begin{center}
\scalebox{0.90}{%
\begin{tabular}{ccr}
\hline\hline
States Li $\rightarrow$ Be$^+$ &   Amplitude (CI) \quad & \quad P$_{f,i}$ (CI)   \\
\hline
\hline
7$^2$H $\rightarrow$ 7$^2$H & 0.430 251 660 & 18.51 \\ 
7$^2$H $\rightarrow$ 8$^2$H & 0.685 081 584 & 46.93 \\
\hline
8$^2$H $\rightarrow$ 7$^2$H & 0.334 387 142 & 11.18 \\
8$^2$H $\rightarrow$ 8$^2$H & 0.192 337 137 &  3.70 \\
\hline
8$^2$I $\rightarrow$ 8$^2$I & 0.382 098 764 & 14.60 \\
\hline
\end{tabular}}
\end{center}
\end{table}

\end{document}